# Color variations of Comet C/2013 UQ4 (Catalina)

Oleksandra Ivanova[1,2*], Evgenij Zubko[3], Gorden Videen[4,5], Michael Mommert[6], Joseph L. Hora[7], Zuzana Seman Krišandová[1], Ján Svoreň[1], Artyom Novichonok[8], Serhii Borysenko[2], Olena Shubina[2]

[1] Astronomical Institute of the Slovak Academy of Sciences, SK-05960 Tatranská Lomnica, Slovak Republic

[*]Corresponding Author. E-mail address: oivanova@ta3.sk

[2] Main Astronomical Observatory of National Academy of Sciences, Kyiv, Ukraine

[3] School of Natural Sciences, Far Eastern Federal University, 8 Sukhanova Street, Vladivostok 690950, Russia

[4] Space Science Institute, 4750 Walnut Street, Suite 205, Boulder, CO 80301, USA

[5] U.S. Army Research Laboratory, 2800 Powder Mill Road, Adelphi, MD 20783, USA

[6] Northern Arizona University, Department of Physics and Astronomy, PO Box 6010, Flagstaff, AZ, 86011, USA

[7] Harvard-Smithsonian Center for Astrophysics, 60 Garden Street, MS65, Cambridge, MA 02138-1516, USA

[8] Petrozavodsk State University, 33 Lenina Street, Petrozavodsk, Russia

**Abstract**

We report observations of color in the inner coma of Comet C/2013 UQ4 (Catalina) with the broadband *B* and *R* filters. We find significant temporal variations of the color slope, ranging from –12.67 ± 8.16% per 0.1 μm up to 35.09 ± 11.70% per 0.1 μm. It is significant that the comet changes color from red to blue over only a two-day period. Such dispersion cannot be characterized with an average color slope. We also observe Comet C/2013 UQ4 (Catalina) in infrared using Spitzer and find no significant $CO/CO_2$ gaseous species in its coma. Therefore, we classify Comet C/2013 UQ4 (Catalina) as a dust-rich comet and attribute the measured color slope to its dust. We analyze the color slope using the model of agglomerated debris particles and conclude that the C/2013 UQ4 coma was chemically heterogeneous, consisting of at least two components. The first component producing the bluest color is consistent with Mg-rich silicates. There are three different options for the second component producing the reddest color. This color is consistent with either Mg-Fe silicates, kerogen type II, or organic matter processed with a low dose of UV radiation.

# 1. Introduction

The spectral properties of sunlight scattered from a comet differ significantly from those of unscattered light. The change is the result of the specific light-scattering properties of dust particles and gases forming the coma, which are wavelength dependent. Therefore, colorimetric measurements may be used to constrain physical and chemical properties of cometary species. What emerges from the photometric measurements of numerous comets is that their color significantly differs from one comet to another (e.g., Jewitt & Meech 1986; Lamy et al. 2011). For instance, all nine comets investigated in Jewitt & Meech (1986) revealed red photometric color, but of differing strengths. The red color is often considered to be a distinctive feature of cometary dust (e.g., Hanner 2003).

However, other studies demonstrate that comets may also have blue color (e.g., Jewitt et al. 1982; Weiler et al. 2003; Lamy et al. 2011; Zubko et al. 2014; Korsun et al., 2010; Ivanova et al. 2013). For example, all four comets whose coma was investigated by Lamy et al. (2011) displayed blue photometric color. Interestingly, Lamy et al. considered their finding with caution, as they suspected that the blue color occurred due to contamination of the light-scattering response with emission from cometary gases, even though no compelling evidence for the presence of such emission was provided. Some comets that were classified with confidence as dusty comets, in which the light-scattering response from their coma is dominated by the refractory materials, also display blue color. For example, Comets C/1995 O1 (Hale-Bopp) and C/1975 V1 (West) were found to be blue in appearance (see, Weiler et al. 2003 and Zubko et al. 2014, respectively). In Comet Hale-Bopp, the color oscillated between red and blue. Other examples of blue color of cometary dust can be found in Jewitt et al. (1982), when comparing the high-resolution spectrum of Comet C/1980 E1 (Bowell) with the solar spectrum; the spectrum of Comet Bowell appears to be almost featureless and, therefore, Jewitt et al. concluded that it is completely devoid of gaseous emission. Finally, it is worth noting the recent *Rosetta* measurements of the color of dust in Comet 67P/Churyumov–Gerasimenko (Cremonese et al. 2016), in which 16 out of 70 individual dust grains detected in the vicinity of the nucleus have blue photometric color. Thus, the blue photometric color also can be a feature of some comets or, at least, some part of their dust.

One cometary feature that is still poorly addressed in the literature concerns color variation of a given comet over time. Observations of comets often are limited and, as a consequence, observational data correspond typically to rare and random epochs. One of the most systematic surveys spanning a time period of nearly five years, was conducted of the color of Comet C/1995

O1 (Hale-Bopp) (Weiler et al. 2003), which revealed short-term variations with a scale of a few days. Interestingly, such variations did not correspond with the heliocentric distance of the comet. However, Comet Hale-Bopp was found to be exceptional in different senses (e.g., Hadamcik & Levasseur-Regourd 2003; Wooden et al. 1999). A recent study of Comet C/2013 A1 (Siding Spring) with the Hubble Space Telescope also demonstrated that the color slope could vary up to two times over a period of a few months (Li et al. 2014). Nevertheless, unlike in the case of Comet Hale-Bopp, the color of Comet Siding Spring remained red throughout all three epochs of observation.

Systematic, long-term studies of the color are still rarely reported in the literature and, therefore, it is difficult to draw statistically reliable conclusions. In this short paper, we report our observations of Comet C/2013 UQ4 (Catalina) that further enriches the knowledge of the variation of cometary color. These observations reveal fast and significant temporal variations. For instance, we found its color changes from red to blue over only a two-day period. This has not been noted in previous studies. We also provide a quantitative analysis of these observations, which constrain the physical and the chemical properties of the dust in Comet Catalina.

## 2. Techniques of observation and data processing

Comet C/2013 UQ4 (Catalina) was discovered on October 23, 2013. It was not active at the time of discovery and, therefore, was not immediately recognized as a comet. However, the images obtained by Novichonok and Prystavski on May 7, 2014 using the iTelescope Observatory (MPEC 2014-K36) located at Siding Spring revealed cometary features and, as a consequence, the object has been classified as a comet. The comet has a retrograde orbit that is characterized with: a = 60.4 AU, e = 0.98, i = 145º, q = 1.1 AU, P ~ 469 years. Assuming optical albedo of the Catalina nucleus being ~4%, Binzel et al. (2013) constrained its size to the range of 7– 23 km. Using the WISE data obtained before this object showed cometary activity, A. K. Mainzer estimated the diameter of the object as 23.5 km based on thermal modeling.[1] The rotational period of C/2013 UQ4 (Catalina) is unknown at this time.

We started observing Comet Catalina on June 2, 2014, and continued until August 12, 2014, near perihelion passage. Photometric observations of the comet were conducted with the *B* and *R* broadband filters. The reduction of the raw data included bias subtraction, dark and flat field corrections, and cleaning cosmic-ray tracks. The morning sky was exposed to provide a flat field correction for the non-uniform sensitivity of the CCD chip. We refer the Reader to Ivanova et al. (2015, 2016) for more details on the reduction technique. The images were obtained during several observing runs at different sites: the 61-cm Telescope at the Skalnate Pleso (AI SAS, Slovakia), iTelescope T9 RCOS 12.5″ Siding Spring (Australia) and iTelescope T17 Planewave 17" CDK Siding Spring (Australia). Basic parameters of the telescopes are given in Table 1. The general view of the comet is presented in Fig. 1.

---

[1] http://echo.jpl.nasa.gov/asteroids/2013UQ4/2013UQ4_planning.html

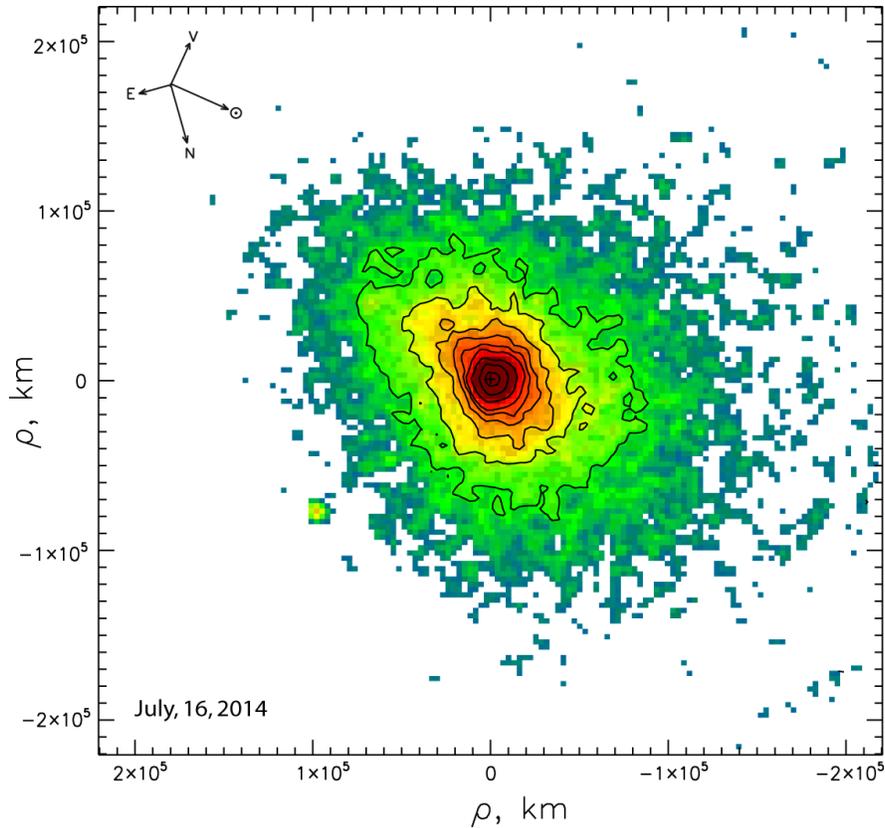

Fig. 1: The image of Comet C/2013 UQ4 (Catalina) observed through a broad band R. Celestial north, east, the motion V and sunward directions are denoted.

Due to its distinctive comet-like orbit, 2013 UQ4 was also observed with the IRAC instrument (Fazio et al. 2004) onboard the Spitzer Space Telescope as part of a search for activity in near-Earth objects in comet-like orbits (Program ID 10109). The target was observed in HDR mode (frame times 1.2 s and 30 s) in IRAC channels 1 (3.6 μm) and 2 (4.5 μm) on 2014-02-26 (pre-perihelion, at a heliocentric distance r=2.22 au), 2014-08-09 (shortly after its perihelion, r=1.22 au), and 2014-09-01 (post-perihelion, r=1.42 au). Observations were reduced, combined in the

moving frame of the target, and analyzed as described in Trilling et al. (2016). A PSF subtraction

| Telescope | Diameter, [m] | CCD | Pixel size, [μm×μm] | Scale[b], [″/pix] | Field of view, [′×′] |
|---|---|---|---|---|---|
| TSP[a] | 0.61 | SBIG-ST-10XME | 9×9 | 1.6 | 19×13 |
| iTelescope T17 Planewave 17" CDK | 0.43 | FLI ProLine PL4710 | 13×13 | 0.92 | 15.5×15.5 |
| iTelescope T9 12.5" RCOS | 0.32 | SBIG ST8 XME | 9×9 | 0.65 | 16.8×11.2 |

technique as described in Mommert et al. 2014 has been applied to study the target's coma.

**Table 1.** Characteristics of the instruments

[a] – TSP- Telescope at the Skalnaté Pleso
[b] – value take into account the applied binning

In the present analysis, we include only the observations conducted on photometric nights. To perform an absolute flux calibration of the comet images, the field stars were used. The stellar magnitudes of the standard stars were taken from the catalogue NOMAD[2] (Ochsenbein et al., 2000; Zacharias et al. 2005). We exploit only non-variable standard stars that came into the field of view. As a result, we obtain the integrated magnitude as a function of the aperture in the observations of the comet.

We use images obtained with the *B* and *R* filters to infer the magnitude of Comet Catalina. The apparent magnitude is determined as follows:

$$m_c(\lambda) = -2.5 \cdot \lg\left[\frac{I_c(\lambda)}{I_s(\lambda)}\right] + m_{st} - 2.5 \cdot \lg P(\lambda) \Delta M , \qquad (1)$$

where $m_{st}$ is the magnitude of the standard star, $I_s$ and $I_c$ are the measured fluxes of the star and the comet, respectively; *P* is the sky transparency that is dependent on the wavelength λ; and Δ*M* is the difference between the cometary and the stellar air masses. Since we are using field stars for the calibration, they are observed at the same airmass and, therefore, we set Δ*M* = 0. Results of observations are summarized in Table 2.

---
[2] http://vizier.u-strasbg.fr/viz-bin/VizieR

**Table 2.** Photometry of the comet C/2013 UQ4 (Catalina).

| Data, UT/ Telescope | [a] Orb. | [b] $r_c$, au | [c] $\Delta$, au | [d] $\alpha$, ° | $m_B$ | $m_R$ | aperture radius | | S′, % per 0.1 μm |
|---|---|---|---|---|---|---|---|---|---|
| | | | | | | | " | km | |
| 2014-06-02, 19:39:45 iTelescope T9 | I | 1.208 | 1.405 | 44.8 | 16.32±0.17 | 15.45±0.24 | 3.2 | 3260.8 | −5.81 ± 18.64 |
| 2014-06-25, 09:53:29 iTelescope T17 | I | 1.095 | 0.655 | 65.5 | 14.87±0.10 | 14.15±0.08 | 7.6 | 3610.4 | −12.67 ± 8.16 |
| 2014-07-15, 00:06:15 TSP | O | 1.091 | 0.359 | 68.6 | 14.14±0.11 | 13.10±0.06 | 12.8 | 3332.8 | 2.03 ± 7.83 |
| 2014-07-16, 20:59:31 TSP | O | 1.096 | 0.398 | 68.1 | 13.77±0.09 | 12.87±0.06 | 12.8 | 3694.8 | −4.43 ± 6.90 |
| 2014-07-18, 21:39:19 TSP | O | 1.102 | 0.450 | 67.3 | 14.46±0.12 | 13.05±0.07 | 9.6 | 3133.2 | 18.88 ± 8.44 |
| 2014-07-19, 21:14:56 TSP | O | 1.105 | 0.478 | 66.8 | 14.43±0.09 | 13.34±0.07 | 9.6 | 3328.1 | 4.34 ± 7.36 |
| 2014-08-01, 20:50:30 TSP | O | 1.166 | 0.901 | 57.1 | 16.77±0.17 | 14.98±0.12 | 3.2 | 2091.1 | 35.09 ± 11.70 |
| 2014-08-12, 09:04:45 iTelescope T17 | O | 1.239 | 1.254 | 48.0 | 16.86±0.13 | 15.46±0.14 | 3.7 | 3365.1 | 18.43 ± 11.99 |

[a] Orbital arc: I is inbound leg of orbit (pre-perihelion); O is outbound leg of orbit (post-perihelion)
[b] $r_c$ is heliocentric distance
[c] $\Delta$ is geocentric distance
[d] $\alpha$ is phase angle of the comet

It is convenient to characterize the cometary color in terms of the so-called *color slope* (A'Hearn et al. 1984). Note, this characteristic sometimes also is referred to as the *reflectivity gradient* (Jewitt and Meech 1986). The color slope can be measured in per cent per 0.1 μm and is defined as follows:

$$S' = \frac{10^{0.4\Delta m} - 1}{10^{0.4\Delta m} + 1} \times \frac{20}{\lambda_2 - \lambda_1}, \qquad (2)$$

where $\Delta m$ stands for the true color index of the comet, i.e., the observed color reduced for the initial solar color. We adapt the color index of the Sun $B - R = 0.996$ from Holmberg et al. (2006). In eq. (2), $\lambda_1$ and $\lambda_2$ correspond with the effective wavelengths of the filters used. In the present work, $\lambda_1 = 0.4353$ μm in the case of the *B* filter ($\Delta\lambda = 0.0781$ μm) and $\lambda_2 = 0.6349$ μm in the case of the *R* filter ($\Delta\lambda = 0.1066$ μm). Although the filters are broadband, their bandwidths are well isolated one from another.

It is important to note that the geocentric distance of the comet varied by a factor of three over the run of observations. However, we measure nearly the same part of the coma, adjusting the angular aperture of the signal integration. In the vast majority of cases, the projected radius of the measured area is squeezed between 3133.2 km and 3694.8 km with only one exception, 2091.1 km corresponding to August 1, 2014. The results of computations of the color slope with eq. (2) are summarized in the last column of Table 2 and, also are shown in Fig. 2 as a function of the phase angle of Comet Catalina.

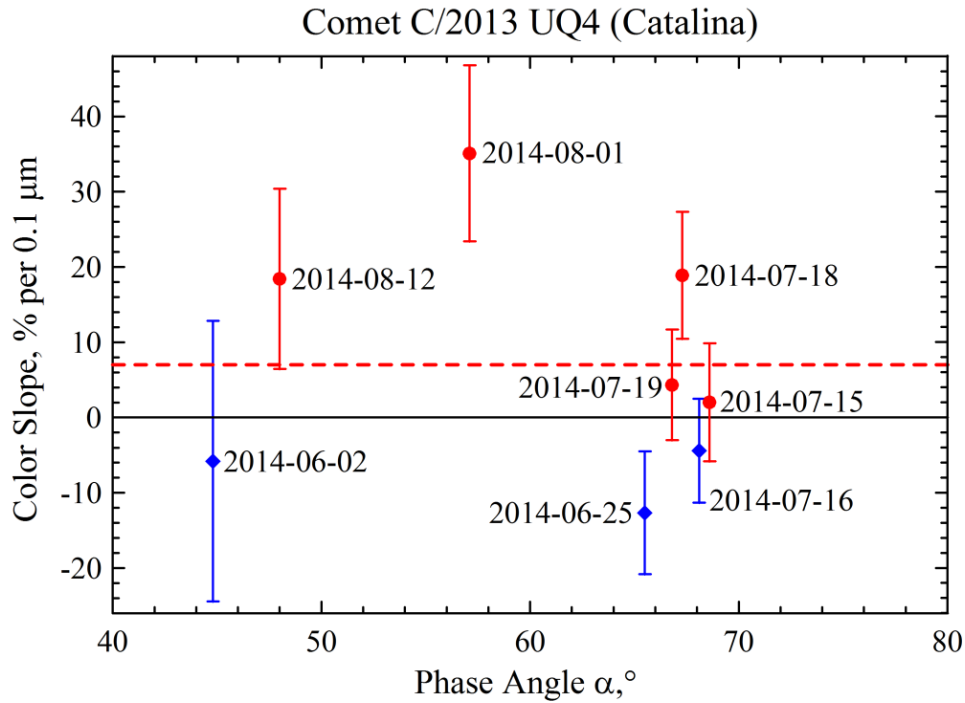

Fig. 2: The color slope as a function of phase angle in Comet C/2013 UQ4 (Catalina). Red dots and blue diamonds show positive and negative values of the color slope, respectively. Every data point is labeled with the corresponding date of observation. The red dashed line shows the average color slope for eight available observations.

## 3. Color slope in Comet Catalina

As one can see in Fig. 2, the color slope $S'$ in Comet Catalina ranges from $(-12.67 \pm 8.16)$% per 0.1 μm up to $(35.09 \pm 11.70)$% per 0.1 μm. We note that the observations were conducted at different phase angles, which can cause some variation of the color (Zubko et al. 2014). However, seven out of eight observations of Comet Catalina fall quite well into the two squeezed ranges of phase angle, $\alpha \approx 44.8° - 48°$ and $\alpha \approx 65.5° - 68.6°$. It is significant that within each of those, the observed variation of the color slope cannot be attributed to the change in the geometry of observations and, therefore, they have to be considered in terms of change of the coma constituents.

What also emerges from Fig. 2 is that the coma population may experience dramatic short-term variations. For instance, during a single day spanning July 18-19, the color slope decreases from $(18.88 \pm 8.44)$% per 0.1 μm to $(4.34 \pm 7.36)$% per 0.1 μm; whereas, over a two-day period spanning July 16-18, the color slope increases from $(-4.43 \pm 6.90)$% per 0.1 μm to $(18.88 \pm 8.44)$% per 0.1 μm. Note, the latter case could also suggest qualitative changes in the Catalina color, from blue to red.

These findings have an important practical implication as they reveal that a sole observation hardly characterizes the color of a comet. Indeed, when characterizing the color of Comet Catalina, we meet a fundamental difficulty due to the large dispersion of the observational results. Namely, in three out of eight observations the coma appears blue and in the remaining five, it appears red. During two epochs, June 2 and July 16, the blue color was detected with high uncertainty that also might suggest a slightly red color of the coma. A similar high uncertainty accompanies the measurements of the red color on July 15 and 19. Overall, with confidence we detect blue color on one night and red color on three other nights. Thus, we face the question: what is the color of Comet Catalina?

The average color slope of Comet Catalina is equal to 6.98% per 0.1 μm that is consistent with other comets (e.g., Hanner 2003). However, this is misleading, since this average color slope does not match any specific observation shown in Fig. 2. While four observations appear consistent with the average color slope within the error bars, the other four observations differ from it considerably. Therefore, in our analysis we do not consider the average color slope in Comet Catalina; instead, we analyze the whole range of the observed color slopes observed.

Before proceeding with the analysis, it is important to address the question of possible contamination of the signal measured with broadband filters with emission from gaseous species in Comet Catalina. Such emission was previously suspected to be responsible for the blue photometric color in some comets (e.g., Lamy et al. 2011). However, in the follow-up analysis we consider that the color variations in Comet Catalina are mainly caused by changes in its dust population. There are two reasons for this. (1) In comets whose blue color was inferred from their high-resolution spectra, it was found that the gaseous emission only slightly (<5%) affects the total signal integrated over a wide range of wavelength (e.g., Jewitt et al. 1982). (2) Comparison of the color index inferred from the emission-free parts in the high-resolution spectrum and directly measured with the broadband *B* and *R* filters reveals remarkably similar results that is virtually the same within the error bars (Weiler et al. 2003). It also is worth noting that a similar trend was found in comparative analyses of the polarimetric response measured in comets with the narrowband continuum filters isolated from the gaseous emission and broadband filters (e.g., Zubko et al. 2014). These findings suggest a generally weak contribution of the gaseous emission onto the total flux measured in comets with the broadband filters. Unfortunately, such validation cannot be directly applied to Comet Catalina because the necessary data have not been reported in the literature yet; and, most likely, they do not exist. Therefore, we can only extrapolate the general trend found in other comets for the Catalina case. The strong emission line of CN gas that can be clearly seen in Cochran et al. (1992) appears at wavelength less than 0.39 um. This is out of the bandpass of the B filter used in the current study and, therefore, it cannot affect our conclusions; whereas, $NH_2$ emission lines indeed may appear within the B filter bandpass. What is important, however, it is that the $NH_2$ lines also lie within the integration range (0.5 um – 0.7 um) considered in Jewitt et al. (1982). For example, they can be clearly seen along with the $C_2$ in Comet Panther (see on middle in Fig. 3 of Jewitt et al. 1982). Despite those emission lines the integration of spectrum reveals their very little relative contribution to the Panther spectrum, less than 1%.

This case reveals very important feature: Gaseous emission lines appearing in cometary spectrum itself do not necessarily affect the signal. What is truly important is the ratio of the fluxes caused by gaseous emission and continuum.

We investigate the possibility of contamination of our optical broadband photometry due to gaseous activity based on our Spitzer observations (see Section 2). While the pre-perihelion observations do not show any kind of activity around the nucleus, the perihelion and post-perihelion data clearly show extended emission around the nucleus. From the PSF-subtracted images, which only show the object's tail and not contributions from the nucleus, we measure the coma flux density in a box that covers the tail in IRAC channels 1 and 2. We place this box as close as possible to the nucleus, avoiding image artifacts that arise from the PSF subtraction process. From our perihelion observations, we derive a channel 2 over channel 1 flux-density ratio of 5.4±1.1 in a box 30,000 km to 65,000 km from the nucleus. Using those data with the shortest frame times obtained as part of our HDR observations, we find a flux density ratio of 4.5±5.5 at a distance of 14,000 km to 30,000 km from the nucleus, barely constraining that ratio. Based on the dust emission model by Reach et al. (2013) (see Figure 3 in that work), these flux ratios agree with emission from dust alone and do not suggest the existence of CO or $CO_2$ in the tail of the comet. In order to verify this assessment, we repeat our analysis on the post-perihelion observations and find a channel 2 over channel 1 flux-density ratio of 5.4±2.6 at a distance of 50,000 km to 80,000 km from the nucleus, which is in high agreement with our perihelion observations. Although our sample distances from the nucleus in these Spitzer data are much larger than our optical observations, we believe that our findings apply to much closer nucleus distances. At the smallest heliocentric distance at which 2013 UQ4 was observed by Spitzer (1.22 au), the lifetime for dissociation by sunlight is 8.6 days for $CO_2$ and 22.4 days for CO (based on Ootsubo et al., 2012, assuming an inverse-square relationship between the lifetime and the heliocentric distance). Assuming an expansion velocity of gas of 0.72 km $sec^{-1}$, and using the relationship provided by Ootsubo et al. (2012) at that heliocentric distance, the dissociation lifetimes translate into distances from the nucleus at which the respective species can exist of 500,000 km for CO2 and 1,400,000 km for CO, both of which are well outside the nucleus distances at which our flux densities were derived. Thus, the *Spitzer* data suggest that the dust contribution dominates over that of $CO/CO_2$ gases in the Catalina coma. We stress, CO and/or $CO_2$ are the second-most abundant gaseous species after $H_2O$ in cometary coma (e.g., Bockelée-Morvan et al. 2004). Furthermore, since $CO/CO_2$ gases well correlate with other gaseous species in comets, one can conclude that Comet Catalina was a dust-rich and gas-poor comet. This suggests that the potential risk of contamination of our observations with the gaseous emissions is extremely low.

## 4. Analysis of observations

We analyze the color in Comet Catalina using model *agglomerated debris particles*. Their shape is highly irregular in appearance (see ten examples in Fig. 3), resembling that of micron-sized craters produced by cometary dust on the *Stardust* space probe during its close encounter with Comet 81P/Wild 2 (see in Hörz et al. 2006) and of interplanetary dust particles.[3] The packing density of constituent material in the agglomerated debris particles is about 0.236; i.e., on average, only 23.6% of the volume of the sphere circumscribed around an agglomerated debris particle is occupied with material. It is significant that under reasonable assumptions about the chemical composition of cometary dust, the bulk material density of agglomerated debris particles (0.35 – 0.83 g/cm$^3$) appears remarkably consistent with what was found *in situ* in cometary dust (0.3 – 3 g/cm$^3$ according to Hörz et al. 2006) and interplanetary dust (0.2 – 2.2 g/cm$^3$ according to Flynn & Sutton 1991). We refer to Section 3 of Zubko et al. (2016) for a detailed comparative analysis of the agglomerated debris particles versus cometary and interplanetary dust, as well as other models of cometary dust particles. The algorithm for particle generation and technique for computation of their light-scattering response also are described there in detail.

---

[3] http://stardust.jpl.nasa.gov/images/science/idp-l.jpg

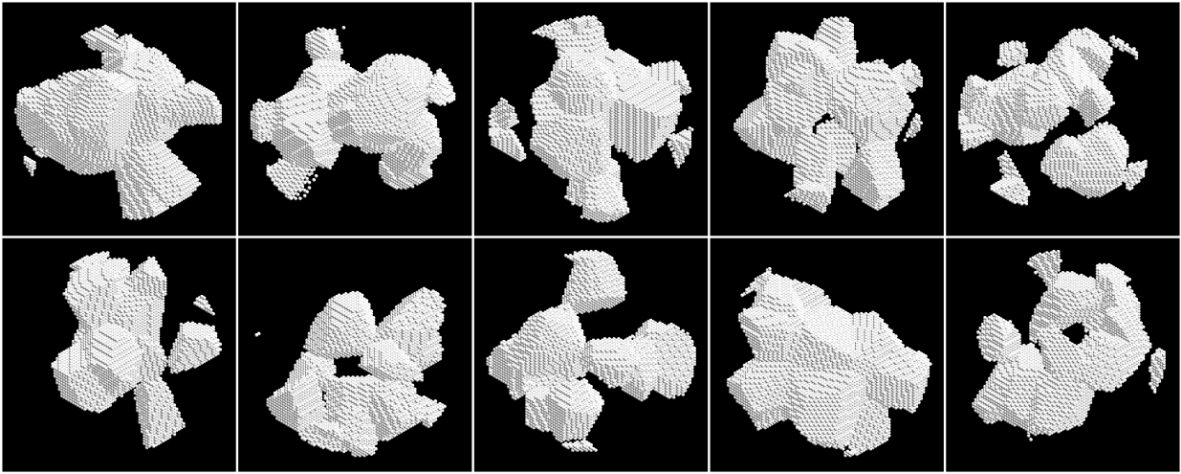

Fig. 3: Ten examples of the agglomerated debris particles

We have calculated optical properties of agglomerated debris particles having 34 different complex refractive indices that are listed on the right in Fig. 4. The vast majority of these are representative of well-known and/or plausible species of comets in the visible. For instance, $m = 1.313 + 0i$ corresponds to water ice (Warren 1984), $m = 2.43 + 0.59i$ to amorphous carbon (Duley 1984), $m = 1.6 + 0.0005i$ and $m = 1.758 + 0.0844i$ to Mg-rich and Mg-Fe silicates, respectively (Dorschner et al. 1995). Refractive indices with Re($m$) = 1.6 and Im($m$) = 0.01 – 0.1 correspond to kerogen type II (Khare et al. 1990). Re($m$) = 1.5 – 1.6 and Im($m$) = 0.05 – 0.15 are consistent with organic materials in a form expected in the diffuse interstellar medium; whereas, the case of $m = 1.855 + 0.45i$ with organics more heavily processed with UV radiation and ion bombardment

(Jenniskens 1993). Refractive indices with greater material absorption, $\text{Im}(m) \sim 1$, are representative of a cosmic carbon analog deduced in Zubko et al. (1996).

Light scattering by submicron and micron-sized particles is dependent on the ratio of the particle radius $r$ to the wavelength of incident radiation $\lambda$, which is quantified with the so-called size parameter $x = 2\pi r/\lambda$. It is clear that the change in $\lambda$ affects $x$ and, as a consequence, the light-scattering response. It is important to emphasize that the effect of the size parameter on light scattering often dominates what could be caused by the wavelength dependence of the refractive index $m$. This holds, for instance, in water ice and Mg-rich silicates (Zubko et al. 2014).

We consider two sets of $x$, $x = 1 – 32$ (50 in the case of $m = 1.313 + 0i$ and 26 in the case of $m = 1.7 + 0i$) and $x \approx 0.07 – 22.3$ (34.9 and 18.2, respectively). If the former set of $x$ is attributed to the effective wavelength of the $B$ filter, $\lambda_B = 0.4353$ μm, then, the later set of $x$ corresponds to the wavelength $\lambda = 0.6234$ μm. It is slightly smaller than the true effective wavelength of the $R$ filter, $\lambda_R = 0.6349$ μm. However, this difference does not affect the color slope because, by definition, the parameter is normalized to the wavelength range.

Both sets of $x$ correspond to the same range of particle radii $r \approx 0.069 – 2.217$ μm (3.464 μm and 1.801 μm in cases $m = 1.313 + 0i$ and $1.7 + 0i$, respectively). We average light-scattering properties of the agglomerated debris particles over this range with a power-law size distribution $r^{-n}$. We note that the *in situ* measurements of comets indeed reveal such a size distribution in the domain of submicron and micron-sized particles with $n = 1.5 – 3$ in Comet 1P/Halley (Mazets et al. 1986) and $n \approx 2.9$ in Comet 81P/Wild 2 (Price et al. 2010). However, *in situ* measurements also suggest significant spatial variations of the power index $n$ over coma (Mazets et al. 1987). There also could be temporal variations of the index $n$ (e.g., Zubko et al. 2011). Therefore, we investigate the color slope over a wide range of the power index $n = 1 – 4$ that embraces all available *in situ* findings to date.

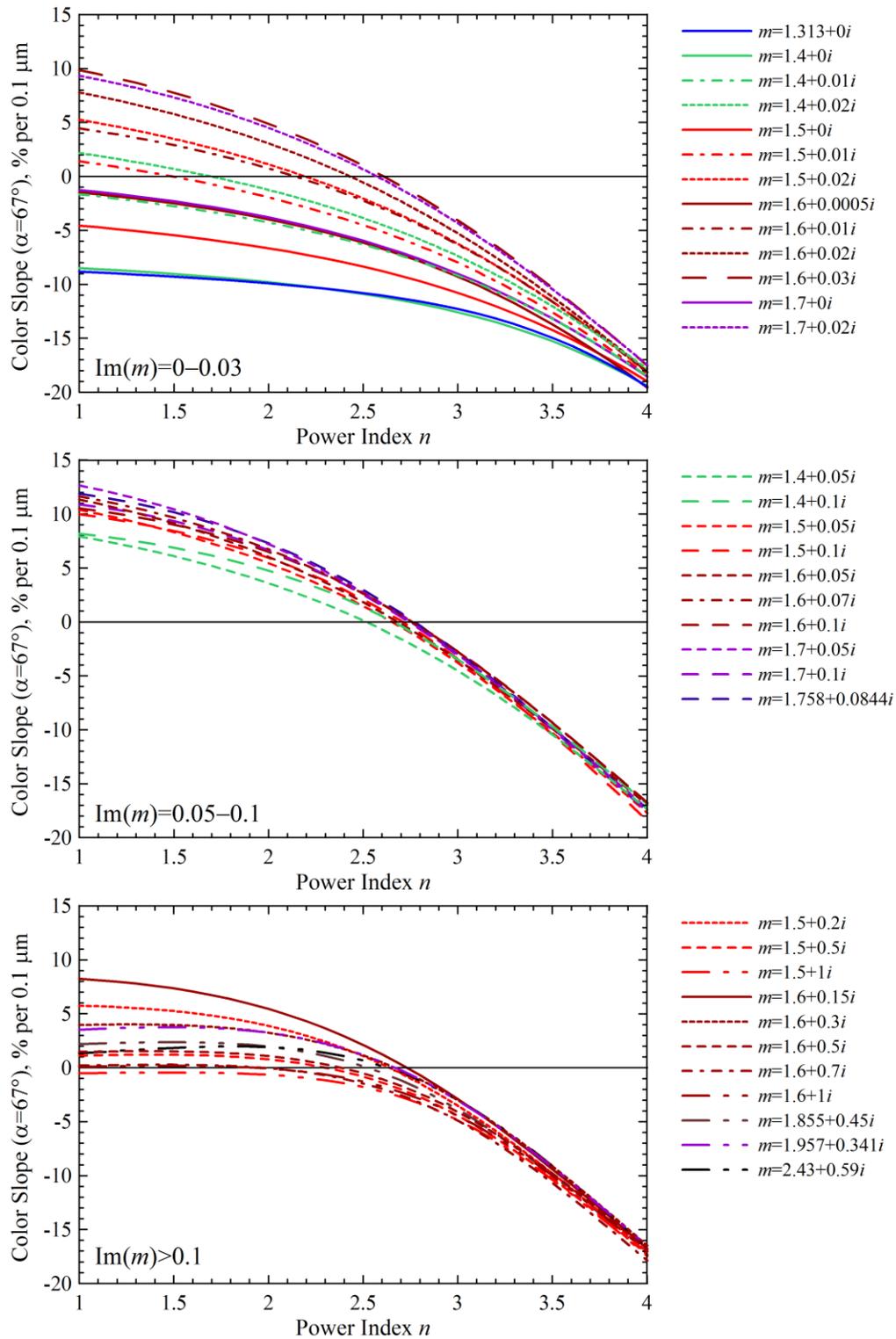

Fig. 4: The color slope as a function of the index *n* in the power-law size distribution of the agglomerated debris particles with different refractive indices *m* (see legend). No wavelength dependence of *m* is assumed.

In Fig. 4 we present the results of our modeling in the assumption of no wavelength dependence of the refractive index $m$. The data correspond to the phase angle $\alpha = 67°$ that is consistent with five out of eight observations demonstrated in Fig. 2. We also repeat such computations at $\alpha = 46°$, which is representative of two other observations, and we found a slight, few per cent per 0.1 µm, non-systematic difference as compared to what is shown in Fig. 4. Therefore, we do not present here the data corresponding to other phase angles. However, we warn the reader against extrapolation of the data presented in Figs. 3 and 4 to other phase angles; in particular, to the case of small phase angles (see, e.g., Fig. 9 of Zubko et al. 2014). Similar results calculated at small phase angle $\alpha = 14°$ can be found in Zubko et al. (2015), which differ by up to 15% per 0.1 µm.

What emerges from Fig. 4 is that no refractive index can reproduce the entire range of the color slope measured in Comet Catalina. The most negative color slope observed on 2014–06–25 can be satisfactorily reproduced with particles having low material absorption. For instance, particles with Re($m$) = 1.5 – 1.6 and Im($m$) = 0 – 0.01 fit $S'$ = –12.67% per 0.1 µm at the power index $n \approx 3.3 – 3.6$. While the given refractive indices appear in good accordance with Mg-rich silicates (Dorschner et al. 1995), a well-known species of comets (e.g., Fomenkova et al. 1992), the power index $n$ is somewhat larger than the upper limit detected in submicron and micron-sized cometary dust *in situ*, i.e. $n = 3$. An assumption on a predominantly carbonaceous composition (e.g., $m = 1.855 + 0.45i$ and $2.43 + 0.59i$) requires a further increase of the power index to $n \approx 3.7 – 3.8$, which is even less compliant with the *in situ* findings. However, when taking into account the uncertainty in measurements of $S'$, one can reduce the power index $n$ in the silicate and carbonaceous particles, adjusting it to the *in situ* measurements of size distribution of cometary dust. Although pure water-ice particles with $m = 1.313 + 0i$ also can fit $S'$ = –12.67% per 0.1 µm at $n \approx 3.1$, their presence in the Catalina coma seems doubtful due to the relatively small solar distance of the comet, $r_h \approx 1.1 – 1.2$ au. In these circumstances, the lifetime of the micron-sized pure water-ice particles does not exceed 1000 seconds (Beer et al. 2006).

Two other negative color slopes detected in Comet Catalina, $S'$ = –5.81% per 0.1 µm on 2014–06–02 and $S'$ = –4.43% per 0.1 µm on 2014–07–16, can be reproduced at lower values of the power index $n$. For instance, the previously mentioned case of Mg-rich silicates with Re($m$) = 1.5 – 1.6 and Im($m$) = 0 – 0.01 can fit $S'$ = –5.81% per 0.1 µm at $n \approx 1.7 – 3$; whereas, $S'$ = –4.43% per 0.1 µm can be attained with $n \approx 1 – 2.8$. These ranges of $n$ closely match *in situ* measurements.

Interestingly, the largest three values of the color slope found in Comet Catalina on 2014–07–18, 2014–08–01, and 2014–08–12 appear in excess of what can be achieved in Fig. 4. This fact makes it necessary to incorporate into our modeling a wavelength dependence of the refractive index $m$. In Fig. 5, we consider the wavelength dependence of the imaginary part of refractive index Im($m$). Namely, we assume that, at the effective wavelength of the $B$ filter, Im($m$) takes on a greater value compared to what in the $R$ filter. Such a trend is found in nearly all types of organic matterial (Jenniskens 1993; Khare et al. 1990; Zubko et al. 1996), in amorphous carbon (Duley 1984), and in Mg-Fe silicates (Dorschner et al. 1995). In the vast majority of cases considered in Fig. 5, for simplicity, we assume that the real part of refractive index Re($m$) remains the same in both filters. This is consistent, for instance, with the measurements of the kerogen type II (Khare et al. 1990) and organics in a form expected in the diffuse interstellar medium (Jenniskens 1993). Nevertheless, we also examine the case of simultaneous changes in Re($m$) and Im($m$).

Fig. 5 demonstrates the effect of an incremental increase of the imaginary part of refractive index in the blue filter $\Delta$Im($m$) on the color slope. From top to bottom, $\Delta$Im($m$) increases from 0.01 up to 0.03. In addition, the bottom panel shows results for $\Delta$Im($m$) = 0.109. In the latter case, both parts of $m$ change simultaneously in accordance with the results obtained by Jenniskens (1993) for organics heavily processed with UV radiation and ion bombardment.

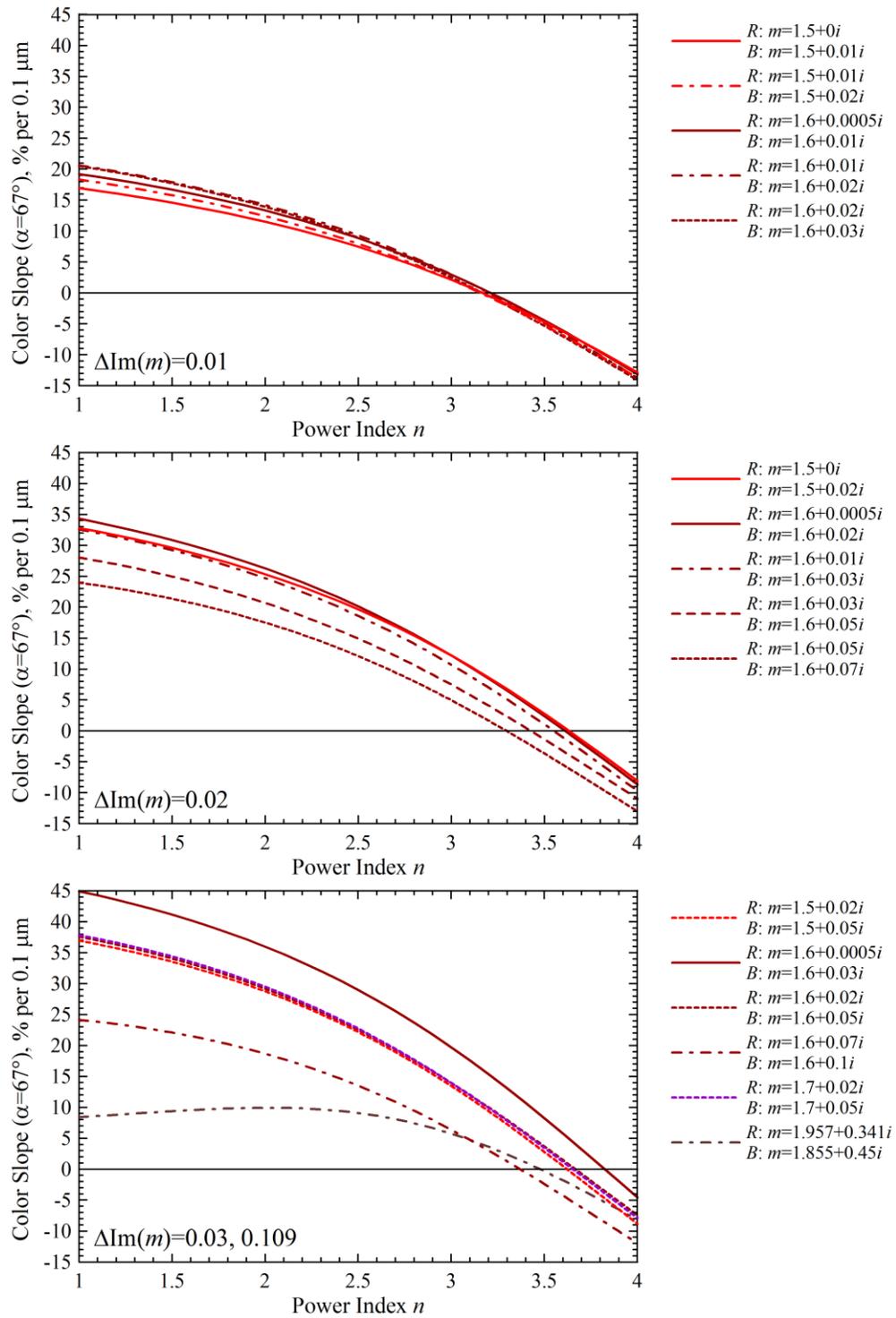

Fig. 5: Similar to Fig. 3, but with the wavelength dependence of the refractive index *m* taken into account (see legend).

As one can see in Fig. 5, the wavelength dependence of Im($m$) significantly increases the reddening compared to that in Fig. 4. Nevertheless, ΔIm($m$) = 0.01 appears insufficient to reproduce the largest value of the color slope measured in Comet Catalina even within its error bars, $S'$ = (35.09 ± 11.70)% per 0.1 µm. Therefore, in order to fit this observation one needs to assume at least ΔIm($m$) = 0.02. This puts a significant constraint on the chemical properties of dust in Comet Catalina. For instance, the pair of $m$ = 1.6 + 0.01$i$ in the $R$ filter and of $m$ = 1.6 + 0.03$i$ in the $B$ filter is representative of kerogen type II (Khare et al. 1990) and some types of low-processed organic materials investigated by Jenniskens (1993). However, this set of refractive indices is capable of fitting the largest color slope only at the bottom limit of the power index, $n$ ≈ 1 that is somewhat lower than one could expect in comets based on *in situ* findings, 1.5 ≤ $n$ ≤ 3. The same also holds for two other cases shown in the middle in Fig. 5 and having Im($m$) = 0 and 0.0005 in the $R$ filter. However, these cases are difficult to attribute to some real materials based on the refractive indices available in the literature to date. We, therefore, leave them out of consideration in the analysis.

The middle panel in Fig. 5 reveals a decreasing effect of ΔIm($m$) = 0.02 on the color slope while the overall material absorption grows. This effect appears even more clear on the bottom panel of Fig. 5 at ΔIm($m$) = 0.03. The largest color slope here, up to 45% per 0.1 µm, is produced by a material with Im($m$) = 0.0005 in the $R$ filter and Im($m$) = 0.03 in the $B$ filter. However, this pair of Im($m$) is difficult to find among known cometary species. Three materials with Im($m$) = 0.02 in the $R$ filter and Im($m$) = 0.05 in the $B$ filter produce a similar, somewhat weaker reddening, with the color slope being up to 38% per 0.1 µm. It is significant that all of them are compliant with known cometary species, such as Mg-Fe silicates (Dorschner et al. 1995) or the organics produced at a low dose of UV radiation (Jenniskens 1993). Interestingly, these materials reproduce the largest color slope in Comet Catalina at the power index $n$ ≈ 1.3 – 1.4; whereas, within the uncertainty of that measurement, the power index can be increased to $n$ ≈ 2.4, which is in good quantitative agreement with the *in situ* findings in comets.

It is worth noting that further increasing the material absorption makes the fit to Comet Catalina impossible. This can be seen on the bottom panel of Fig. 5. We especially draw attention to the pair of refractive indices of $m$ = 1.957 + 0.341$i$ in the $R$ filter and $m$ = 1.855 + 0.45$i$ in the $B$ filter. This closely matches what was measured by Jenniskens (1993) in organic materials heavily processed by UV radiation and ion bombardment. Despite a very high increment of the imaginary part of refractive index $\Delta \text{Im}(m) = 0.109$, the resulting reddening does not exceed 10% per 0.1 μm. This is not consistent with at least two out of five measurements of the positive color slope in Comet Catalina.

Finally, we discuss a mechanism governing the temporal variation of the color observed in Comet Catalina. In general, the two simplest options that may satisfactorily explain the observations are as follows. The first one suggests a chemically heterogeneous coma in Comet Catalina, consisting of at least two types of material: (1) Mg-rich silicate that produces the bluest color in the comet, and (2) Mg-Fe silicates, kerogen type II, and/or organic matter processed with a low dose of UV radiation that all may produce the reddest color in the comet. Such a difference in chemical composition of dust particles could be explained in terms of their origin from different active areas on the Catalina surface. All intermediate values of the color slope can result from different volume ratios of the end members. Taking into account the uncertainty existing in the colorimetric measurements of Comet Catalina, both components may obey the same size distribution with a power index $n \approx 2.4$. However, it also is possible for some variation to occur between these components.

The second scenario suggests that the coma is optically homogeneous. It might consist of Mg-Fe silicates, kerogen type II, and/or organic matter processed with a low dose of UV radiation. As in the first scenario, each of these materials is capable of producing the reddest color measured in Comet Catalina. However, the refractive indices of these materials are somewhat similar in the visible, and we cannot discriminate one material from the other with confidence. Therefore, the Catalina coma could be chemically homogeneous. Within the second scenario, the reddest color in Catalina is achieved at $n = 1.3 - 1.4$; or, if we take into account the uncertainty of the measurement, at $n \approx 2.4$. Then, the other color slopes in Comet Catalina correspond to larger values of the power index $n$, up to $n = 4.2 - 4.3$ on the epoch of the bluest color. We notice that such very high values of $n$ are hardly consistent with the range detected *in situ*. Interestingly, unlike the case of the reddest color, taking into account the uncertainty in the measurement of the bluest color only slightly relaxes the power index to $n = 3.8 - 3.9$, that still is in excess of the *in situ* findings. It is worth noting that the high values of $n$ also would imply very high positive polarization in Comet Catalina at side scattering. With a power index $n = 3.8 - 3.9$ and phase angle $\alpha \sim 70°$, particles with $\text{Im}(m) = 0.02 - 0.05$ produce a degree of linear polarization $P \geq 50\%$ (Zubko et al. 2013). Unfortunately, polarimetric observations of Comet Catalina have not been reported in the literature to date; therefore, we cannot thoroughly examine the second scenario.

However, from what is known in other comets, $P \geq 50\%$ would be a very unusual polarimetric response for a comet at $\alpha \sim 70°$ (e.g., Chernova et al. 1993). We conclude, therefore, that the second scenario is likely inconsistent with our expectations based on *in situ* findings and ground-based observations of other comets. Thus, the first scenario is seemingly more realistic.

## 5. Conclusions

We have measured the color slope in the inner coma (< 4000 km) of Comet Catalina and found rapid variations on a scale of one-two days. Comet Catalina is not a unique case in this sense. For instance, short-term variations in the color slope were previously reported for Comet C/1995 O1 (Hale-Bopp) (Weiler et al. 2003), and long-term variations were detected recently in Comet C/2013 A1 (Siding Spring) (Li et al. 2014). What makes the case of Comet Catalina exceptional is the qualitative changes between the blue color and the red color during a few days. These three cases suggest that a single measurement may hardly characterize the color of a comet. However, the vast majority of photometric observations of comets reported in the literature correspond to a sole epoch (e.g., Jewitt & Meech 1986; Lamy et al. 2011). Such observations should be considered, therefore, with caution because, it is not clear how well they actually characterize comets.

We compute the average color slope in Comet Catalina over eight available observations. We found, however, that it cannot adequately characterize the color of Comet Catalina due to the large dispersion of the results obtained on different epochs. Therefore, we analyze the whole range of values of the color slope in Comet Catalina. Using the model of irregularly shaped agglomerated debris particles, we retrieve physical and chemical properties of dust in Comet Catalina on the epochs when it revealed its reddest and bluest colors. What emerges from this study is that the corresponding extreme values of the color slope are unlikely to be reproduced with the same material. This leads us to an important conclusion on chemical heterogeneity of the Catalina coma that there are at least two components in the coma contributing significantly to the optical response. The first component producing the bluest color is consistent with Mg-rich silicates. There are three different options for the second component producing the reddest color. It could be either Mg-Fe silicates, kerogen type II, or organic matter processed with a low dose of UV radiation, as well as a combination of those three species. However, all intermediate values of the color slope in Comet Catalina can be explained in terms of different volume ratios of these two principle components. Our modeling also suggests that the dust particles in both components obey a power-law size distribution $r^{-n}$ with the index $n$ being consistent with the *in situ* findings in comets.

**Acknowledgements**

O. Ivanova thanks for financial support the SASPRO Programme. The research leading to these results has received funding from the People Programme (Marie Curie Actions) European Union's Seventh Framework Programme under REA grant agreement No. 609427. Research has been further co-funded by the Slovak Academy of Sciences grant VEGA 2/0032/14. This research has made use of the VizieR catalogue access tool, CDS, Strasbourg, France. Grateful acknowledgment is made to Dr. M.S.P. Kelley, who made an announcement in Cometary Science News about searching spectral data for the comet. The authors thank the anonymous reviewer for valuable comments on the manuscript.